\documentclass[conference]{IEEEtran}

\usepackage{amsmath,amssymb} \interdisplaylinepenalty=2500
\usepackage{color}
\usepackage{subfigure}
\usepackage{epsfig}
\usepackage{graphicx}
\usepackage{dsfont}
\usepackage{verbatim}
\usepackage{amsthm}
\usepackage{setspace}

\newtheorem{definition}{Definition}
\newtheorem{theorem}{Theorem}

\newtheorem{claim}{Claim}

\newenvironment{remark}{\textit{Remark: }}{}

\def\qed{\endIEEEproof}

\newcommand{\calB}{\mathcal{B}}

\newcommand{\calD}{\mathcal{T}}
\newcommand{\calE}{\mathcal{E}}

\newcommand{\calG}{\mathcal{G}}

\newcommand{\calS}{\mathcal{S}}
\newcommand{\calT}{\mathcal{T}}

\newcommand{\calV}{\mathcal{V}}

\newcommand{\bfY}{\mathbf{Y}}

\newcommand{\F}{\mathds{F}}
\newcommand{\Fq}{\mathds{F}_{q}}

\newcommand{\Fbm}{\mathds{F}_{2^{m}}}

\newcommand{\NoE}{E}
\newcommand{\NoS}{S}
\newcommand{\Bl}{n}
\newcommand{\lfs}{m}
\newcommand{\mess}{M}
\newcommand{\Cp}{C}

\newcommand{\Tran}{T}
\newcommand{\TranE}{\hat{T}}

\newcommand{\Basis}{B}
\newcommand{\Ma}{M_1}
\newcommand{\Mb}{M_2}

\newcommand{\disXY}{d_{\hat{T}}(X,Y)}
\newcommand{\disYX}{d_{\hat{T}}(Y,X)}
\newcommand{\disYZ}{d_{\hat{T}}(Y,Z)}
\newcommand{\disXZ}{d_{\hat{T}}(X,Z)}

\begin{document}

\title{Binary Error Correcting Network Codes}

\author{
\centering
\begin{tabular}{ccc}
Qiwen Wang & Sidharth Jaggi & Shuo-Yen Robert Li
\\
&
{\small Department of Information Engineering}
&
\\
&
{\small The Chinese University of Hong Kong}
&
\end{tabular}
}

\maketitle

\begin{abstract}
We consider network coding for networks experiencing {\it worst-case bit-flip} errors, and argue that this is a reasonable model for highly dynamic wireless network transmissions. We demonstrate that in this setup prior network error-correcting schemes (\cite{SonYC:06, Kotter.Kschischang2008}) can be arbitrarily far from achieving the optimal network throughput. We propose a new metric for errors under this model. Using this metric, we prove a new {\it Hamming-type} upper bound on the network capacity. We also show a commensurate lower bound based on {\it GV-type} codes that can be used for error-correction. 
The codes used to attain the lower bound are non-coherent (do not require prior knowledge of network topology). 
The end-to-end nature of our design enables our codes to be overlaid on classical distributed random linear network codes~\cite{Ho++2003}. Further, we free internal nodes from having to implement potentially computationally intensive link-by-link error-correction.
\end{abstract}

\IEEEpeerreviewmaketitle

\section{Introduction}
A source Alice wishes to transmit information to a receiver Bob over a network with ``noisy" links. Such a communication problem faces several challenges.

The primary challenge we consider is that in highly dynamic (wireless) environments the noise levels on each link might vary significantly across time, and hence be hard to estimate well.
This issue of variable noise levels exacerbates at least two other challenges that had been considered settled by prior work.

One, since noise exists in the network, network coding might be dangerous.
This is because all nodes mix information, so even a small
number of bit-flips in transmitted packets may end up corrupting all the information flowing in the network, causing decoding errors. Prior designs for network error-correcting codes exist (for {\it e.g.}~\cite{SonYC:06, Kotter.Kschischang2008}) but as we shall see they are ineffective against bit-flips in a highly dynamic noise setting. In particular, one line of work ({\it e.g.}~\cite{Cai.Yeung2002, Jaggi++2007:Byzantine, Kotter.Kschischang2008}) treats even a single bit-flip in a packet as corresponding to the entire packet being corrupted, and hence results in rates that are too pessimistic -- the fundamental problem is that the codes are defined over ``large alphabets", and hence are poor at dealing with bit-flip errors. Another line of work ({\it e.g.}~\cite{SonYC:06}) overlays network coding on link-by-link error correction, but requires accurate foreknowledge of the noise levels on each link to have good performance.


Two, in dynamic settings, the coding operations of nodes in the network may be unknown {\it a priori}. Under the bit-flip error-model we consider, however, the ``transform-estimation" strategy of Ho {\it et al.}~\cite{Ho++2003} does not work, since any headers pre-specified for this use can also end up being corrupted.

In this work we consider simultaneously the {\it reliability} and {\it universality} issues for random linear network coding. Namely, we design end-to-end distributed schemes that allow reliable
network communications in the presence of ``worst-case" network noise, wherein the erroneous bits can be arbitrarily distributed in different network packets with only the constraint that the total number of bit-flips is bounded from above. Internal network nodes just do linear network coding. Error-correction is only carried out at the receiver(s), which also estimates the linear transform imposed on the source's data by the network\footnote{As is common in coding theory, the upper and lower bounds on error-correction we prove also directly lead to corresponding bounds on error-detection -- for brevity we omit discussing error-detection in this work.}.

As noted above, our codes are robust to a wide variety of channel conditions -- whether the noise bit-flips are evenly distributed among all packets, or even adversarially concentrated among just a few packets, our codes can detect and correct errors up to a network-wide bound on the total number of errors.
Na\"ive implementations of prior codes (for instance, of link-by-link error-correcting codes~\cite{SonYC:06}) that try to correct for worst-case network conditions may result in network codes with much lower rates (see the example in Section~\ref{ex:toy} below). Thus the naturally occurring diversity of network conditions works in our favour rather than against us.

Also, even though our codes correct {\it binary} errors rather than errors over larger symbol fields as in prior work, the end-to-end nature of our design enables our codes to be overlaid on classical linear network codes over finite fields (for instance, the random linear network codes of Ho {\it et al}~\cite{Ho++2003}). Further, we free internal nodes from having to implement potentially computationally intensive link-by-link error-correction.

The main tool used to prove our results is a {\it transform metric} that may be of independent interest.
It is structurally similar to the rank-metric used by Silva {\it et al.}~\cite{Kotter.Kschischang2008}, but has important differences that give our codes the power of universal robustness against binary noise (as opposed to the packet-based noise considered in~\cite{Kotter.Kschischang2008},~\cite{Jaggi++2007:Byzantine} and~\cite{SonYC:06}).

\section{Model}
\subsection{Network model}
We model our network by a directed acyclic multigraph\footnote{Our model also allows non-interfering broadcast links in a wireless network to be modeled via a directed hypergraph -- for ease of notation we restrict ourselves to just graphs.}, denoted by $\calG=(\calV,\calE)$, where $\calV$ denotes the set of nodes and $\calE$ denotes the set of edges. A single source node $s\in\calV$ and a set of sinks $\calD\subseteq\calV$ are pre-specified in $\calV$. We denote $|\calE|$ and $|\calD|$, respectively the number of edges and sinks in the network, by $\NoE$ and $\NoS$.
A directed edge $e$ leading from node $u$ to node $v$ can be represented by the vector $(u,v)$, where $u$ is called the {\it tail of $e$} and $v$ is called the {\it head of $e$}. In this case $e$ is called an {\it outgoing edge of $u$} and an {\it incoming edge of $v$}.

The capacity of each edge is one packet -- an length-$\Bl$ vector over a finite field $\Fbm$ -- here
$\Bl$ and $\lfs$ are design parameters to be specified later.
Multiple edges between two nodes are allowed -- this allows us to model links with different capacities.\footnote{By appropriate buffering and splitting edges into multiple edges, any network can be approximated into such a network with unit capacity edges.}
As defined in~\cite{ACLY00}, the {\it network (multicast) capacity}, denoted $\Cp$, is the minimum over all sinks $t \in \calD$ of the mincut of $\calG$ from the source $s$ to the sink $t$. Without loss of generality, we assume there are $\Cp$ edges outgoing from $s$ and incoming edges to $t$ for all sinks $t \in \calD$.
\footnote{In cases where the number of outgoing edges from $s$ (or the number of incoming edges to $t$) is not $\Cp$, we can add a {\it source super-node} (or {\it sink super-node}) with $\Cp$ noiseless {\it edges} connecting to the original source (or sink) of the network. The change in the number of edges and probability of error on each edge are small compared to those of the original network, so our analysis essentially still applies.}


\subsection{Code model}
The source node $s$ wants to {\it multicast} a message $\mess$ to each sinks $t \in \calD$.
To simplify notation, we consider henceforth just a single sink -- our analysis can be directly extended to the multi-sink case. All logarithms in this work are to the base $2$, and we use $H(p)$ to denote the {\it binary entropy function} $-p\log p -(1-p)\log(1-p)$.

\noindent {\bf Random linear network coding:}
All internal nodes in the network perform {\it random linear network coding}~\cite{Ho++2003} over a finite field $\Fbm$. Specifically, each internal node takes uniformly random linear combinations of each incoming packet to generate outgoing packets.
That is, let $e'$ and $e$ index incoming and outgoing edges from a node $v$. The {\it linear coding coefficient from $e'$ to $e$} is denoted by $f_{e',e}\in\Fq$. Let ${\mathbf Y_{e}}$ denote the packet (length-$\Bl$ vector over $\Fbm$) transmitted on the edge $e$. Then $\bfY_{e}=\sum f_{e',e}\bfY_{e'}$, where the summation is over all edges $e'$ incoming to the node $v$, and all arithmetic is performed over the finite field $\Fbm$.

\noindent {\bf Mapping between $\F_2$ and $\Fbm$:}
The noise considered in this work is binary in nature. Hence, to preserve the linear relationships between inputs and outputs of the network,
we use the mappings given in Lemma~$1$ in~\cite{JagEHM:04}. These map addition and multiplication over $\Fbm$ to corresponding (vector/matrix) operations over $\F_2$.
More specifically, a bijection is defined from each symbol (from $\Fbm$) of each packet transmitted on each edge, to a corresponding length-$\lfs$ bit-vector. For ease of notation henceforth, for each edge $e$ and each $i \in \{1,\ldots,\Bl\}$, we use ${\mathbf Y}_e$ and ${\mathbf Y}_e(i)$ solely to denote respectively the length-$\Bl\lfs$ and length-$\lfs$ binary vectors resulting from the bijection operating on packets and their $i$th symbols, rather than the original analogues over $\Fbm$ traversing that edge $e$. Separately, each {\it linear coding coefficient} $f_{e',e} $ at each node is mapped via a homomorphism to a specific $\lfs \times \lfs$ binary matrix $F_{e',e} $. The linear mixing at each node is then taken over the binary field -- each length-$\lfs$ binary vector ${\mathbf Y}_{e'}(i)$ (corresponding to the binary mapping of the $i$th symbol of the packet ${\mathbf Y}_{e'}$ over the field $\Fbm$) equals $\sum F_{e',e}{\mathbf Y}_{e'}(i)$. It is shown in~\cite{JagEHM:04} that an isomorphism exists between the binary linear operations defined above, and the original linear network code. In what follows, depending on the context, we use the homomorphism to switch between the scalar (over $\Fbm$) and matrix (over $\F_2$) forms of the network codes' linear coding coefficients, and the isomorphism to switch between the scalar (over $\Fbm$) and vector (over $\F_2$) forms of each symbol in each packet.

\noindent {\bf Noise:}
We consider {``worst-case noise"} in this work, wherein an arbitrary number of bit-flips can happen in any transmitted packet, subject to the constraint that no more that a fraction of $p$ bits over all transmitted packets are flipped.
The {\it noise matrix} $Z$ is
an $\NoE \lfs\times \Bl$ binary matrix with at most $p\NoE\lfs \Bl$ nonzero entries which can be arbitrarily distributed. In particular, the $\lfs(i-1)+1$ through the $\lfs i$ rows of $Z$ represent the bit flips in the $i$th packet ${\mathbf Y}_{e_i}$ transmitted over the network. If the $(k\lfs + j)$th bit of the length-$\lfs\Bl$ binary vector is flipped (that is, the $j$th bit of the $k$th symbol over $\Fbm$ in ${\mathbf Y}_{e_i}$ is flipped), then the $(\lfs(i-1)+j,k)$ bit in $Z$ equals $1$, else it equals $0$.
Thus the noise matrix $Z$ represents the {\it noise pattern} of the network.
To model the noise as part of the linear transform imposed by the network, we add an artificial super-node $s'$ connected to all the edges in the network, injecting noise into each packet transmitted on each edge in the network according to entries of the noise matrix $Z$.


\noindent {\bf Source:}
The source has a set of $2^{R\lfs\Bl}$ messages $\{\mess\}$ it wishes to communicate to each sink, where $R$ is the {\it rate} of the source. Corresponding to each message $\mess$ it generates a {\it codeword} $X(\mess)$ using the encoders specified in Section~\ref{subsec:GV} (to make notation easier we usually do not explicitly reference the parameter $\mess$ and instead refer simply to $X$). This $X$ is represented by a $\Cp \times \Bl$ matrix over $\Fbm$, or alternatively a $\Cp\lfs \times \Bl$ matrix over $\F_2$. Each row of this matrix corresponds to a packet transmitted over a distinct edge leaving the source.


\noindent {\bf Receiver(s):}
Each sink $t$ receives a batch of $\Cp$ packets. Similarly to the source, it organizes the received packets into a matrix $Y$, which can be equivalently viewed as a $\Cp \times \Bl$ matrix over $\Fbm$ or a $\Cp\lfs \times \Bl$ binary matrix.
Each sink $t$ decodes the message $\hat{\mess}$ from the received matrix $Y$ using the decoders specified in Section~\ref{subsec:GV}.

%
\noindent {\bf Transfer matrix and Impulse response matrix:}
Having defined the linear coding coefficients of internal nodes, the packets transmitted on the incoming edges of each sink $t$ can inductively be calculated as linear combinations of the packets on the outgoing edges of $s$. We denote the $\Cp \times \Cp$ {\it transfer matrix from the outgoing edges of $s$ to the incoming edges of $t$ by $\Tran$}, over the finite field $\Fbm$. Alternatively, using the homomorphism described above, $\Tran$ may be viewed as as ${\Cp \lfs \times \Cp \lfs}$ binary matrix.

We similarly define $\TranE$ to be the {\it impulse response matrix}, which is the transfer matrix from a imaginary source $s'$--who injects errors into all edges--to the sink $t$. Note that $\Tran$ is a sub-matrix of $\TranE$, composed specifically of the $\Cp$ columns of $\Tran$ corresponding to the $\Cp$ outgoing edges of $s$.

In this work we require that every $\Cp \times \Cp$ sub-matrix of $\TranE$ is invertible.
As noted in, for instance,~\cite{Koetter.Medard2003, Ho++2003}
this happens with high probability for random linear network codes. Alternatively, deterministic designs of network error-correcting codes~\cite{Cai.Yeung2002} also have this property.

Using the above definitions the network can thus be abstracted by the equation~\eqref{received} below as a {\it worst-case binary-error network channel}.
\begin{equation}\label{received}
  Y = \Tran X + \TranE Z.
\end{equation}
Similar equations have been considered before (for instance in~\cite{Cai.Yeung2002, Jaggi++2007:Byzantine, Kotter.Kschischang2008}) -- the key difference in this work is that we are interested in $Z$ matrices which are fundamentally best defined over the binary field, and hence, when needed, transform the other matrices in \eqref{received} also into binary matrices.

%

%

\noindent{\bf Performance of code:} The source encoders and channel decoders specified in Section~\ref{subsec:GV} together comprise {\it worst-case binary-error-correcting network codes}.
A {\it good} worst-case binary-error-correction network channel code has the property that, for all messages $\mess$, and noise patterns $Z$ with at most $p\NoE\lfs\Bl$ bit-flips, $\hat{\mess} = \mess$.
A rate $R$ is said to be {\it achievable} for the worst-case binary-error channel if, for all sufficiently large $\Bl$, there exists a good code with rate $R$.

\subsection{Toy Example}~\label{ex:toy}
\begin{figure}\label{toy}
    \centering
        \includegraphics[width=0.3\textwidth]{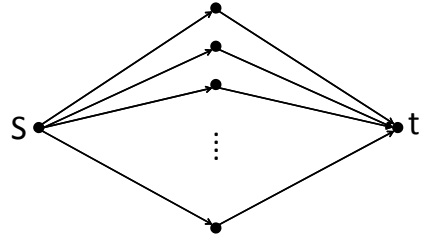}
    \caption{A network with $\Cp$ parallel paths from the source to the destination. Each internal node performs random linear network coding.}
\end{figure}

We demonstrate via an example that in networks with worst-case bit-errors, prior schemes have inferior performance compared to our scheme. In Figure~\ref{toy} the network has $\Cp$ paths (with a total of $2\Cp$ links that might experience worst-case bit-flip errors).

\noindent {\it Benchmark 1:} If link-by-link error-correction\footnote{Since interior nodes might perform network coding, na\"ive implementations of end-to-end error-correcting codes are not straightforward -- indeed -- that is the primary goal of our constructions.} is applied as in~\cite{SonYC:06}, {\it every} link is then required to be able to correct $2\Cp p \Bl$ {\it worst-case} bit-flip errors (since all the bit-errors may be concentrated in any single link). Using GV codes (\cite{Gil52,Var57}) a rate of $1-H(4 \Cp p)$ is achievable on each link, and hence
the overall rate scales as $\Cp(1-H(4 \Cp p))$. As $\Cp$ increases without bound, the throughput thus actually goes to zero. The primary reason is that every link has to prepare for the worst case number of bit-flips aggregated over the entire network, but in large networks, the total number of bit-flips in the worst-case might be too much for any single link to be able to tolerate.

\noindent {\it Benchmark 2:} Consider now a more sophisticated scheme, combining link-by-link error correction with end-to-end error-correction as in~\cite{Kotter.Kschischang2008}. Suppose each link can correct $\frac{2\Bl \Cp p}{k}$ worst-case bit-flips, where $k$ is a parameter to be determined such that the rate is optimized. Then at most $k$ links will fail. Overlaying an end-to-end network error-correcting code as in~\cite{Kotter.Kschischang2008} with link-by-link error-correcting codes such as GV codes (effectively leading to a concatenation-type scheme) leads to an overall rate of $(\Cp-2k)(1-H(\frac{4\Cp p}{k}))$. For large $\Cp$, this is better than the previous benchmark scheme since interior nodes no longer attempt to correct {\it all} worst-case errors and hence can operate at higher rates -- the end-to-end code corrects the errors on those links that do experience errors. Nonetheless, as we observe below, our scheme still outperforms this scheme, since concatenation-type schemes in general have lower rates than single-layer schemes.

\noindent {\it Our scheme:} The rate achieved by our scheme (as demonstrated in Section~\ref{subsec:GV}) is at least $\Cp(1-2H(2p))$. As can be verified, this rate is higher than either of the benchmark schemes.


\section{Transform Metric}\label{subsec:metric}
We first define a ``natural" distance function between binary matrices $\Ma$ and $\Mb$ related as $\Ma=\Mb + \Basis Z$ for some matrices $\Basis$ and $Z$.

Let $\Ma$ and $\Mb$ be arbitrary $a \times b$ binary matrices. Let $\Basis $ be a given $a \times c$ matrix with full column rank. Let $\Ma(i)$ and $\Mb(i)$ denote respectively the $i$th columns of $\Ma$ and $\Mb$. We define $d_\Basis(\Ma,\Mb)$, the {\it transform distance between $\Ma$ and $\Mb$ in terms of $\Basis$}, as follows.
\begin{definition}
\label{def:dis}
Let $\delta(i)$ denote the minimal number of columns of $\Basis$ that need to be added to $\Ma(i)$ to obtain $\Mb(i)$. Then the transform distance $d_\Basis(\Ma,\Mb)$ equals $\sum_{i=1}^b \delta(i)$.
\end{definition}
\medskip

%

\begin{figure}
    \centering
        \includegraphics[width=0.4\textwidth]{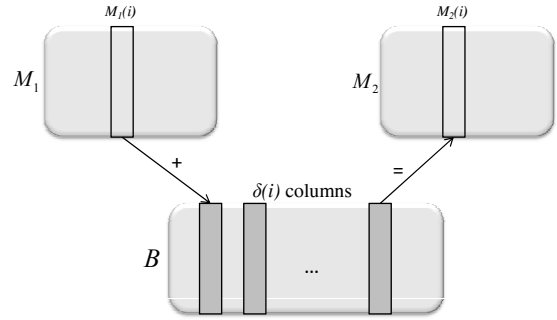}
    \caption{Transform metric: the minimal number of columns of $\Basis$ that need to be added to $\Ma(i)$ to obtain $\Mb(i)$ is $\delta(i)$.}
\end{figure}

%
%
%

\section{Main results}

In this section we present our main results. In Theorem~\ref{thm:Hammingbd1} in Subsection~\ref{subsec:Ham} we present an upper bound on the rates of communication achievable by any code over networks that have ``worst-case" bit-flip errors. Our bounding technique is motivated by the corresponding Hamming bound technique in classical coding theory~\cite{Hamming:50} -- the main challenge lies in deriving good {\it lower} bounds for the ``volumes of spheres" in
the channel model and corresponding metric defined in Section~\ref{subsec:metric}.

In Subsection~\ref{subsec:GV} we
discuss schemes that achieve ``good" rates of communication over networks that have ``worst-case" bit-flip errors. We present three schemes motivated by the well-known Gilbert-Varshamov (GV) bound from classical coding theory~\cite{Gil52,Var57} -- again, the challenge lies in deriving good {\it upper} bounds on the volumes of  spheres in the metric we define. Theorem~\ref{thm:GVbdCo} considers the {\it coherent} scenario, {\it i.e.}, when the linear coding coefficients in the network (or at least the transfer matrix $\Tran$ and the impulse response matrix $\TranE$) are known in advance to the receiver. We use this setting primarily for exposition, since the proof is somewhat simpler than the proof for the {\it non-coherent} setting, when no advance information about the topology of the network, the linear coding coefficients used, or $\Tran$ or $\TranE$ is known in advance to the receiver. In Theorem~\ref{thm:GVbdInco} we are able to demonstrate that essentially the same rates as in Theorem~\ref{thm:GVbdCo} are still achievable, albeit with an rate-loss that is asymptotically negligible in the block-length $\Bl$. 

As we see below, the functional forms of both the Hamming-type upper bounds and the GV-type lower bounds we derive are structurally very similar to those of the classical Hamming and GV bounds.


\subsection{Hamming-type bound}\label{subsec:Ham}
\medskip
\begin{theorem}
\label{thm:Hammingbd1} For all $p$ less than ${\Cp}/({2\NoE\lfs})$ an upper bound on the achievable rate of any code over the worst-case binary-error channel is $1-H(p)(\frac{\NoE}{\Cp})$.
\end{theorem}
\noindent {\it Proof:}
Since each transmitted codeword $X$ is a $\Cp \lfs \times \Bl$ binary matrix, the number of possible choices of $X$ is at most $2^{\Cp \lfs \Bl}$.
But suppose $X$ is transmitted, by the definitions of the worst-case bit-error channel, the received $Y$ lies in the radius-$p\NoE\lfs\Bl$ ball (in the transform metric)
$\calB_{\TranE}(\Tran X,p\NoE\lfs\Bl)$ defined as $\{Y|d_{\TranE}(\Tran X,Y)\leq p\NoE\lfs\Bl\}$. For the message corresponding to $X$ to be uniquely decodable, it is necessary that the balls $\calB_{\TranE}(\Tran X,p\NoE\lfs\Bl)$ be non-intersecting for each $X$ chosen to be in the codebook.
Hence to get an upper bound on the number of codewords that can be chosen, we need to derive a lower bound of the volume of $\calB_{\TranE}(\Tran X,p\NoE\lfs\Bl)$. Recall that $Y$ equals $\Tran X+\TranE Z$. Hence we need to bound from below the number of distinct values of $\TranE Z$ for $Z$ with at most  $p\NoE \lfs \Bl$ ones.

We consider the case that $Z$ has exactly $p\NoE \lfs \Bl$ ones that are equally distributed among columns of $Z$ -- hence every column of $Z$ has $p\NoE \lfs$ ones in it.
We now show that, in the worst case, every such distinct matrix $Z$ results in distinct $\TranE Z$.
Suppose not -- in that case there exist distinct $Z$ and $Z'$ with $p\NoE \lfs$ ones in each columns of both matrix such that $\TranE Z$ equals $\TranE Z'$, {\it i.e.}, $\TranE (Z-Z')$ equals the zero matrix. In particular, for at least some column of $Z$ and $Z'$, say $Z(i)$ and $Z'(i)$, it must be the case that $\TranE (Z(i) - Z'(i))$ equals $0$. But by assumption each column of both $Z$ and $Z'$ has less that $p\NoE \lfs < \Cp/2$ ones, and hence $Z(i)-Z'(i)$ has less than $\Cp$ ones in it.

We now view $\TranE$ and $Z$ as matrices over $\Fbm$. From the argument above, $Z(i)-Z'(i)$ has less than $\Cp$ non-zero elements over $\Fbm$ in it (since an element over $\Fbm$ is zero if and only if each of the $\lfs$ bits in its binary representation is zero). Hence $\TranE (Z(i)-Z'(i))$ is a linear combination over $\Fbm$ of strictly less than $\Cp$ columns of $\TranE$.
But as to the matrix $\TranE$ viewed over $\Fbm$, since we are deriving a worst-case upper bound, we can also require that every $\Cp \times \Cp$ sub-matrix of $\TranE$ is invertible (as noted before 
this happens with high probability for random linear network codes). Hence $\TranE (Z(i)-Z'(i))$ cannot equal the zero vector, which leads to a contradiction.


Hence the number of distinct values for
$\TranE Z$ is at least the number of distinct values for $Z$ with at most $p\NoE \lfs$ ones in each column. This equals is at least ${{\NoE\lfs}\choose{p\NoE\lfs}}^\Bl$, which by Stirling's approximation~\cite{CoverT:91} is at least $2^{H(p)\NoE\lfs\Bl-\log(\NoE\lfs+1)}$. The total number of $\Cp \lfs \times \Bl$ binary matrices is $2^{\Cp \lfs \Bl}$.
Thus an upper bound on the size of any codebook for the worst-case binary-error channel is
\begin{equation*}
    \frac{2^{\Cp \lfs \Bl}}{2^{\NoE \lfs \Bl H(p)-\log(\NoE\lfs+1)}} = 2^{(1-H(p)\frac{\NoE}{\Cp} +\frac{\log(\NoE\lfs+1)}{\Cp \lfs \Bl})\Cp \lfs \Bl},
\end{equation*}
which, asymptotically in $\Bl$, gives the Hamming-type upper bound on the rate of any code as $1-H(p)\frac{\NoE}{\Cp}$.
\hfill $\Box$

\subsection{Gilbert-Varshamov-type bounds}\label{subsec:GV}

\subsubsection{Coherent GV-type network codes}
We first discuss the case when the network transfer matrix $\Tran$ and impulse response matrix $\TranE$ are known in advance.

\noindent {\it Codebook design:} Initialize the set $\calS$ as the set of all binary $\Cp \lfs \times \Bl$ matrices.
Choose a uniformly random $\Cp \lfs\times \Bl$ binary matrix $X$ as the first codeword. Eliminate from $\calS$ all matrices in the radius-$2p\NoE\lfs\Bl$ ball (in the transform metric)
$\calB_{\TranE}(\Tran X,2p\NoE\lfs\Bl)$.
Then choose a matrix $Y'$ uniformly at random in the remaining set and choose $X'=\Tran^{-1}Y'$ as the second codeword. Now, further eliminate all matrices in the radius-$2p\NoE\lfs\Bl$ ball
$\calB_{\TranE}(\Tran X',2p\NoE\lfs\Bl)$ from $\calS$, choose a random $Y'$ from the remaining set, and choose the third codeword $X''$ as $X''=\Tran^{-1}Y''$.
Repeat this procedure until the set $\calS$ is empty.

\medskip
\begin{theorem}
\label{thm:GVbdCo} Coherent GV-type network codes achieve a rate of
at least $1-H(2p)\frac{\NoE}{\Cp}$.
\end{theorem}
\noindent {\it Proof:}
For this theorem, we need an upper bound on $\calB_{\TranE}(\Tran X,2p\NoE\lfs\Bl)$ (rather than a lower bound on $\calB_{\TranE}(\Tran X,p\NoE\lfs\Bl)$ as in Theorem~\ref{thm:Hammingbd1}).
Recall that $Y=\Tran X+ \TranE Z$
The number of different $Y$, or equivalently, different $\TranE Z$, can be bounded from above by the number of different $Z$. This equals $\displaystyle\sum\limits_{i=0}^{2p\NoE \lfs \Bl} {{\NoE \lfs \Bl}\choose{i}}$. The dominant term this summation is when $i$ equals $2p\NoE \lfs \Bl$. Hence the summation can be bounded from above by $(2p\NoE \lfs \Bl+1){{\NoE \lfs \Bl}\choose{2p\NoE \lfs \Bl}}$. By Stirling's approximation~\cite{CoverT:91} this is at most $(2p\NoE \lfs \Bl+1)2^{H(2p)\NoE\lfs\Bl}$.

Thus a lower bound on the size of the codebook for coherent GV-type
\begin{equation*}
    \frac{2^{\Cp \lfs \Bl}}{(2p\NoE \lfs \Bl+1)2^{H(2p)\NoE \lfs \Bl}} = 2^{(1-H(2p)\frac{\NoE}{\Cp}-\frac{\log(2p\NoE \lfs \Bl+1)}{\Bl})\Cp \lfs \Bl},
\end{equation*}
which, asymptotically in $\Bl$, gives the rate of coherent GV-type bound network codes as $1-H(2p)\frac{\NoE}{\Cp}$.
\hfill $\Box$

\medskip

\subsubsection{Non-coherent GV-type network codes}
The assumption that $\Tran$ and $\TranE$ are known in advance to the receiver is often unrealistic, since the random linear coding coefficients in the network are usually chosen on the fly.
Hence we now consider the non-coherent setting, wherein $\Tran$ and $\TranE$ are not known in advance. We demonstrate that despite this lack of information the same rates as in Theorem~\ref{thm:GVbdCo} are achievable in the non-coherent setting.

The number of all possible $\TranE$ is at most by $2^{\Cp\NoE\lfs}$ since $\TranE$ is a ${\Cp\times \NoE}$ matrix $ \Fbm$ -- the crucial observation is that this number is independent of the block-length $\Bl$.
Hence in the non-coherent GV setting, we consider {\it all} possible values of $\TranE$ (and hence $\Tran$, since it comprises of a specific subset of $\Cp$ columns of $\TranE$).

\noindent {\it Codebook design:} Initialize the set $\calS$ as the set of all binary $\Cp \lfs \times \Bl$ matrices.
Choose a uniformly random $\Cp \lfs\times \Bl$ binary matrix $X$ as the first codeword. For each
$\Cp\times \NoE$ matrix $\TranE$ (over the field $\Fbm$),
eliminate from $\calS$ all matrices in the radius-$2p\NoE\lfs\Bl$ ball (in the transform metric)
$\calB_{\TranE}(\Tran X,2p\NoE\lfs\Bl)$.
Then choose a matrix $Y'$ uniformly at random in the remaining set and choose $X'=\Tran^{-1}Y'$ as the second codeword. Now, further eliminate all matrices in the radius-$2p\NoE\lfs\Bl$ ball
$\calB_{\TranE}(\Tran X',2p\NoE\lfs\Bl)$ from $\calS$, choose a random $Y'$ from the remaining set, and choose the third codeword $X''$ as $X''=\Tran^{-1}Y''$.
Repeat this procedure until the set $\calS$ is empty.

\medskip
\begin{theorem}
\label{thm:GVbdInco} Non-coherent GV-type network codes achieve a rate of
at least $1-H(2p)\frac{\NoE}{\Cp}$.
\end{theorem}
\noindent {\it Proof:}
The crucial difference with the proof of Theorem~\ref{thm:GVbdCo} is in the process of choosing codewords -- at each stage of the codeword elimination process, at most $2^{\Cp\NoE\lfs}|\calB_{\TranE}(\Tran X',2p\NoE\lfs\Bl)|$ potential codewords are eliminated (rather than $|\calB_{\TranE}(\Tran X',2p\NoE\lfs\Bl)|$ potential codewords as in Theorem~\ref{thm:GVbdCo}). Hence the number of potential codewords that can be chosen in the codebook is at least
$$\frac{2^{\Cp \lfs \Bl}}{2^{\Cp\NoE\lfs}(2p\NoE \lfs \Bl+1)2^{H(2p)\NoE \lfs \Bl}}$$
which equals$$
2^{(1-H(2p)\frac{\NoE}{\Cp}-\frac{(\log(2p\NoE \lfs \Bl+1)+\NoE)}{\Bl})\Cp \lfs \Bl}.
$$ As can be verified, asymptotically in $\Bl$ this leads to the same rate of $1-H(2p)\frac{\NoE}{\Cp}$ as in Theorem~\ref{thm:GVbdCo}.
\hfill $\Box$

{\it Note:} Our proposed codes via concatenation schemes so that their encoding and decoding complexity grows only polynomially in the block-length (albeit exponentially in network parameters).

\subsection{Scale of Parameters}\label{subsec:scale}
We now investigate the regime of $p$ wherein our results are meaningful.

\begin{claim}\label{scalep}
For all $p$ less than $\min(\frac{\Cp}{2\NoE\lfs}, \frac{1}{2^{\lfs+1}})$ the Hamming-type bounds and GV-type hold.
\end{claim}
\noindent {\it Proof:} The Hamming-type bound in Theorem~\ref{thm:Hammingbd1} requires $p\NoE\lfs < \frac{\Cp}{2}$.

For the GV-type bound in Theorems~\ref{thm:GVbdCo} and~\ref{thm:GVbdInco} to give non-negative rates, $H(2p)\frac{\NoE}{\Cp} < 1$. Hence when $p$ is very small,
\begin{align}
H(2p)\frac{\NoE}{\Cp}
&\rightarrow 2p(\log{1/(2p)})\frac{\NoE}{\Cp} \label{p0}\\
&< \frac{\Cp}{\NoE\lfs}(\log{1/(2p)})\frac{\NoE}{\Cp} \label{p1}= \frac{-\log{2p}}{\lfs}\\
&<1 \label{p2}
\end{align}
where \eqref{p0} follows from the limiting behaviour of the binary entropy function for small $p$,
\eqref{p1} is because $p < \Cp/(2\NoE\lfs)$ (our first condition), and \eqref{p2} is because $p < 2^{-(\lfs +1)}$ (our second condition).

\section{Conclusion}
In this work we investigate upper and lower bounds for the performance of end-to-end error-correcting codes for worst-case binary errors. This model is appropriate for highly dynamic wireless networks, wherein the noise-levels on individual links might be hard to accurately estimate. We demonstrate significantly better performance for our proposed schemes, compared to prior benchmark schemes. 


\section*{Acknowledgment}
The work was partially supported by a grant from the University Grants Committee of the Hong Kong Special Administrative Region, China (Project No. AoE/E-02/08), the CERG grant 412207, and Project MMT-p7-11 of the Shun Hing Institute of Advanced Engineering, The Chinese University of Hong Kong.




%

\bibliographystyle{plain}
\bibliography{sid_necc}

\begin{thebibliography}{10}

\bibitem{ACLY00}
R.~Ahlswede, N.~Cai, S.-Y.~R. Li, and R.~W. Yeung.
\newblock Network information flow.
\newblock {\em IEEE Transactions on Information Theory}, 46(4):1204--1216,
  2000.

\bibitem{Cai.Yeung2002}
Ning Cai and R.~W. Yeung.
\newblock Network coding and error correction.
\newblock In {\em Proc. 2002 IEEE Inform. Theory Workshop}, pages 119--122,
  Bangalore, India, October 20--25, 2002.

\bibitem{CoverT:91}
T.~Cover and J.~Thomas.
\newblock {\em Elements of Information Theory}.
\newblock John Wiley and Sons, 1991.

\bibitem{Gil52}
E.~N. Gilbert.
\newblock A comparison of signalling alphabets.
\newblock {\em Bell Systems Technical Journal}, 31:504--522, 1952.

\bibitem{Hamming:50}
R.~W. Hamming.
\newblock Error detecting and error correcting codes.
\newblock {\em Bell System Technical Journal}, 29:147--160, 1950.

\bibitem{Ho++2003}
T.~Ho, R.~K{\"o}tter, M.~M{\'e}dard, D.~R. Karger, and M.~Effros.
\newblock The benefits of coding over routing in a randomized setting.
\newblock In {\em Proc. IEEE Int. Symp. Information Theory}, page 442,
  Yokohama, Japan, June 29--July 4, 2003.

\bibitem{JagEHM:04}
S.~Jaggi, M.~Effros, T.~Ho, and M.~M\'{e}dard.
\newblock On linear network coding.
\newblock In {\em Proceedings of 42nd Annual Allerton Conference on
  Communication, Control, and Computing}, Monticello, IL, 2004.

\bibitem{Jaggi++2007:Byzantine}
S.~Jaggi, M.~Langberg, S.~Katti, T.~Ho, D.~Katabi, and M.~M{\'e}dard.
\newblock Resilient network coding in the presence of {B}yzantine adversaries.
\newblock In {\em Proc. 26th IEEE Int. Conf. on Computer Commun.}, pages
  616--624, Anchorage, AK, May 2007.

\bibitem{Koetter.Medard2003}
R.~K{\"o}tter and M.~M\'{e}dard.
\newblock An algebraic approach to network coding.
\newblock {\em IEEE Transactions on Information Theory}, 11(5):782--795,
  October 2003.

\bibitem{SonYC:06}
R.~W.~Yeung, L.~Song and N.~Cai.
\newblock A separation theorem for single-source network coding.
\newblock {\em IEEE Transactions on Information Theory}, 52(5):1861--1871,
  2006.

\bibitem{Kotter.Kschischang2008}
D.~Silva, F.~R. Kschischang and R.~K{\"o}tter.
\newblock A rank-metric approach to error control in random network coding.
\newblock {\em IEEE Transactions on Information Theory}, 54(9):3951--3967, Sep.
  2008.
  
\bibitem{Var57}
R.~R. Varshamov.
\newblock Estimate of the number of signals in error correcting codes.
\newblock {\em Dokl. Acad. Nauk}, 117:739--741, 1957.

\end{thebibliography}

\end{document}